\documentstyle[epsf,rotate]{mn}

\title[An Irradiation Effect in DN Gem]
      {An Irradiation Effect in Nova DN Gem 1912 and the Significance of
the Period Gap for Classical Novae}

\author[A. Retter, E.M. Leibowitz \& T. Naylor]
   {A.~Retter$^{1,2}$,
  E.M.~Leibowitz$^2$
   and T. Naylor$^1$\\
 $^1$Dept. of Physics, Keele University, Keele, Staffordshire, 
ST5 5BG, U.K.; ar@astro.keele.ac.uk; timn@astro.keele.ac.uk\\
 $^2$School of Physics and Astronomy and the Wise Observatory,
Raymond and Beverly Sackler Faculty of Exact Sciences,\\
Tel-Aviv University, Tel Aviv, 69978, Israel; elia@wise.tau.ac.il\\}

\date{Accepted 1999 April 1. Received ???; in original form ???}

\begin{document}

\maketitle

\begin{abstract}

Continuous CCD photometry of the classical nova DN Gem during 52 nights 
in the years 1992-98 reveals a modulation with a period 0.127844 d. The 
semi-amplitude is about 0.03 mag. The stability of the variation suggests 
that it is the orbital period of the binary system. This interpretation 
makes DN Gem the fourth nova inside the cataclysmic variable (CV) period 
gap, as defined by Diaz \& Bruch (1997), and it bolsters the idea that 
there is no period gap for classical novae. However, the number of known 
nova periods is still too small to establish this idea statistically. 
We eliminate several possible mechanisms for the variation, and propose 
that the modulation is driven by an irradiation effect. We find that 
model light curves of an irradiated secondary star, fit the data well. 
The inclination angle of the system is restricted by this model to 
$10^{\circ}$ $\la$ i $\la$ 65$^{\circ}$. We also refine a previous 
estimate of the distance to the binary system, and find d=1.6$\pm$0.6 kpc.

\end{abstract}


\section{Introduction - History of observations on DN Gem}


Nova DN Geminorum is an old classical nova, which erupted in 1912, 
and reached a peak magnitude of $m_{V}\approx 3.5$. The visual light 
curve of the nova a few months after the outburst was characterized 
by two maxima, and by strong oscillations. The nova was recognized 
as a fast one, with $t_{2V}\approx 16d$, and $t_{3V}\approx 37d$ 
(McLaughlin 1960). 

The spectrum of the nova at quiescence shows a strong continuum, with
weak emission lines (Hummason 1938; Williams 1983). Warner (1986) deduced 
an inclination angle of about $50^{\circ}$ from a comparison of the
emission line widths of the nova with a few other classical novae, 
whose inclination angles are known from eclipses, assuming that the 
lines emanate from an accretion disc. Duerbeck, Lemke and 
Willerding (1979) deduced a distance of d=450$\pm70$ pc, 
$A_{V}=0.27\pm0.13$ and $M_{V(max)}=-5.3\pm0.5$ from interstellar lines 
in the nova spectrum.

Robinson and Nather (1977) failed in their search for rapid oscillations 
in DN Gem and nine more classical novae. Retter and Leibowitz (1996), 
however, reported the detection of a sinusoidal variation with a period 
of 0.12785$\pm$0.00005 d and a peak-to-trough amplitude of about 0.06 mag. 
In this work, we present the results of the photometry of Nova Geminorum, 
and discuss in detail the possible mechanisms that can generate the light 
modulation.  Nova DN Gem 1912 is the fourth nova whose period is close 
to the upper edge of the period gap distribution of CVs. In Section 4.7 
we discuss the implications of this fact.

\section{Observations}

We observed Nova DN Gem during 14 nights in the years 1992-93
through an $R$ filter, in the $I$ filter during 34 nights in 1995-97, and
continuously switched between the $I$ and $B$ filters during four nights in
1998.  Table 1 presents a summary of the observation schedule. The
photometry was carried out with the 1-m telescope at the Wise
Observatory, with the Tektronix 1K CCD camera.  Our $R$, $V$ and $B$ 
filters are the standard Cousins filters. The $I$ filter is pseudo-Johnson,
i.e. its red end is determined by the CCD response. The typical 
exposures times were 3-4 m in the $I$ and $R$ bands and 7 m in the $B$ band. 
We note that most of our observations in 1995-8 were obtained during 
bright phases of the moon.

In addition, one snapshot through each of the $B$,$V$,$R$ and $I$ filters was 
taken during each of the nights of 1996 October, 4th and 1997 November,
27th, together with a series of exposures of nearby standard stars. 
Unfortunately, the photometric solutions were poor during both occasions. 
The magnitudes of Nova Gem 1912, measured during the latter 
night are: $m_{B}=15.8$, $m_{V}$=16.0, $m_{R}$=15.7, $m_{I}$=15.4.
Error is about 0.2 mag globally.

\begin{table}
\caption{The observations time table}
\begin{tabular}{@{}ccccc@{}}

UT    & Time of Start  &Run Time&Points&Filter/s\\
Date  &     (HJD)      & (hours)&number&       \\
\\

230192& 2448645.198    &    6.8 & 37   &  R     \\
240192& 2448646.213    &    0.5 & 4    &  R     \\
250192& 2448647.186    &    9.3 & 53   &  R     \\
230492& 2448735.238    &    0.5 & 6    &  R     \\
240492& 2448736.233    &    1.6 & 17   &  R     \\
250492& 2448737.228    &    2.3 & 23   &  R     \\

280193& 2449016.490    &    1.3 & 14   &  R     \\
110293& 2449030.208    &    7.6 & 39   &  R     \\
120293& 2449031.204    &    0.4 & 3    &  R     \\
130293& 2449032.194    &    1.4 & 8    &  R     \\
140293& 2449033.196    &    7.8 & 39   &  R     \\
020493& 2449080.247    &    2.3 & 10   &  R     \\
030493& 2449081.233    &    3.6 & 19   &  R     \\
090493& 2449087.241    &    3.0 & 21   &  R     \\

111295& 2450063.442    &    0.3 & 8    &  I     \\
121295& 2450064.475    &    3.2 & 85   &  I     \\
131295& 2450065.492    &    2.7 & 46   &  I     \\

020196& 2450085.416    &    4.8 & 83   &  I     \\
030196& 2450086.507    &    3.6 & 57   &  I     \\
090196& 2450092.398    &    5.8 & 106  &  I     \\
100196& 2450093.423    &    2.5 & 25   &  I     \\
130196& 2450096.411    &    4.5 & 84   &  I     \\
140196& 2450097.522    &    2.8 & 53   &  I     \\
300196& 2450113.269    &    7.6 & 46   &  I     \\
030296& 2450117.192    &    9.3 & 124  &  I     \\
050296& 2450119.189    &    6.4 & 87   &  I     \\
080396& 2450151.234    &    2.7 & 36   &  I     \\
090396& 2450152.195    &    2.8 & 37   &  I     \\
110396& 2450154.189    &    0.6 & 9    &  I     \\
290396& 2450172.214    &    5.2 & 52   &  I     \\
310396& 2450174.209    &    5.3 & 76   &  I     \\

041096& 2450361.202    &    0.3 &1,1,1,1&I,R,V,B \\

231196& 2450411.493    &    3.2 & 54   &  I     \\
151296& 2450433.431    &    3.8 & 55   &  I     \\
201296& 2450438.375    &    4.9 & 70   &  I     \\
221296& 2450440.310    &    6.9 & 100  &  I     \\

150197& 2450464.251    &    1.6 & 24   &  I     \\
170197& 2450466.194    &    6.8 & 92   &  I     \\
250297& 2450505.274    &    3.5 & 50   &  I     \\
260297& 2450506.188    &    3.0 & 43   &  I     \\
270297& 2450507.199    &    7.5 & 94   &  I     \\
280297& 2450508.197    &    7.6 & 107  &  I     \\
010397& 2450509.295    &    3.3 & 7    &  I     \\
260397& 2450534.196    &    4.2 & 55   &  I     \\
270397& 2450535.216    &    4.3 & 36   &  ?       \\
290397& 2450537.206    &    4.7 & 61   &  I     \\
300397& 2450538.202    &    5.6 & 78   &  I     \\
210497& 2450560.216    &    3.5 & 48   &  I     \\
230497& 2450562.259    &    1.7 & 21   &  I     \\

271197& 2450780.303    &    0.3 &1,1,1,1&I,R,V,B \\

090198& 2450823.253    &    8.3 & 47,45&  I,B   \\
100198& 2450824.279    &    4.0 &   8  &  I       \\
130198& 2450827.509    &    0.6 & 3,3  &  I,B   \\
140198& 2450828.255    &    1.3 & 6,7  &  I,B   \\

\end{tabular}
\end{table}

Aperture photometric measurements were performed using the DAOPHOT program
(Stetson 1987). Instrumental magnitudes of the nova, as well as of 3 to 20 
reference stars, depending on each image quality were obtained from the 
frames. An internally consistent series of nova magnitudes was obtained by 
using the Wise Observatory reduction program DAOSTAT (Netzer et al. 1996).

We did not consider the points of 1997 March 27 in our analysis,
because it seems that a wrong filter was accidentally used. Thus, the
number of remaining frames obtained in each filter on our programme is:
2039 (I), 295 (R), 2 (V) and 57 (B).
 
\section{Data Analysis}

A variation at $P\approx 3$ hr and a full amplitude of about 5 - 10$\%$ 
can be seen by simple visual inspection of the light curve in a few 
of our best nights. In addition, the nightly mean varied by a few tenths 
mag from night to night with $\sigma\approx 0.09$ mag - more than three 
times larger than the mean error for a single data point. In Fig. 1 the 
observations in the $I$ band during 1995-8 are plotted. The points 
represent the raw data, while the circles display nightly means.

\begin{figure}

\centerline{\epsfxsize=3.0in\epsfbox{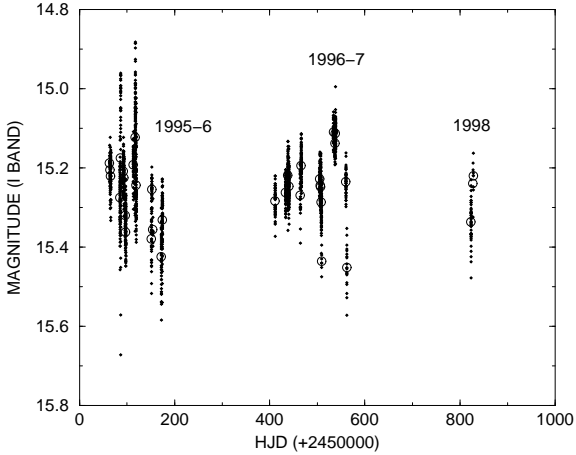}}

\caption{All 1995-8 data points ($I$ band) are presented.  The empty
circles display the mean of each night. There is a variation of the
order of a few tens per cent between adjacent nights.}

\end{figure}

The power spectrum of the 1995-8 data obtained through the $I$ filter is
shown in Fig. 2. The de-trending was carried out by subtracting the mean
from each night. The highest peak in the graph corresponds to the
periodicity of 0.127844$\pm$0.000001 d. This period appears also in the
power spectrum of each of the years 1992, 1993, 1995-6 winter and
1996-7 winter separately. The peak is many $\sigma$ above the noise
level. The fact that it is detected even in the light curves of most of
our long-duration nights makes it also highly significant. The other
peaks seen in the graph around the highest peak are identified as
aliases. The group of peaks at the left hand side of the diagram
corresponds to periodicities that are longer than the typical interval
of observations in each night, and might reflect the noise added to our
data by the presence of the moonlight. A consistent search for a
second real periodicity in the light curve with the various techniques
discussed by Retter, Leibowitz \& Ofek (1997) was not fruitful.

\begin{figure}

\centerline{\epsfxsize=3.0in\epsfbox{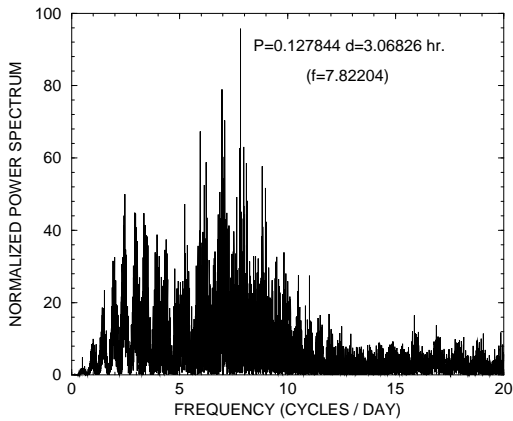}}

\caption{The power spectrum of the 1995-8 points ($I$ band). The highest 
peak at the frequency 7.82204 d$^{-1}$, marked as P, corresponds to the 
period 0.127844 d.}

\end{figure}

In Fig. 3 we present the mean light curve of the 1995-8 observations.
We omitted from the data the nights in which the duration of the
observations was shorter than one full cycle. The light curve shows a
symmetrical sinusoidal variation. The first harmonic fit to the data
yields a semi amplitude of $0.028\pm0.005$. The error corresponds to
a 99\% confidence level, and it was calculated by a sample of 1000 Bootstrap
simulations (Efron \& Tibshirani 1993).

The semi amplitude of the periodicity in the 1992-3 data is very
similar to this value in the 1995-8 data - $0.031\pm0.008$, and the
shape of the mean light curve is a sinusoidal, too. It seems that there
is no systematic difference in the amplitude of the variation in the
two bands ($I$ and $R$). However, the amplitude of the period in the $B$
band, which is derived only from a single night (1998 January, 9th) 
is consistent with zero. Taking into account the interval length of
the observations during this night (8.3 h), and the fact that the
variation is clearly detected in the $I$ band data of the same night, 
we believe that this result is significant.

\begin{figure}

\centerline{\epsfxsize=3.0in\epsfbox{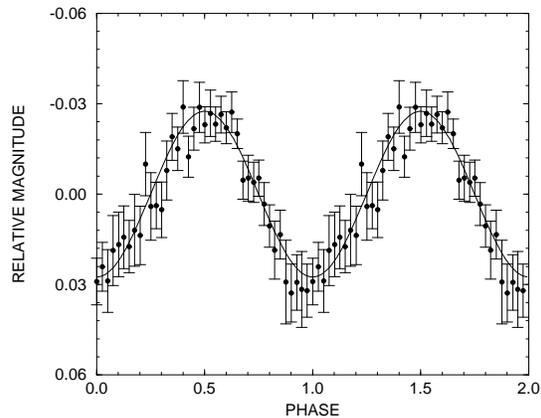}}

\caption{The $I$ filter light curve of the nights in 1995-8, folded on
to the 3.1-h period and binned into 20 equal bins.}

\end{figure}

The best fitted ephemeris of the periodicity is:\\
\\
T${min}$ = HJD 2450823.2866 + 0.127844 E.\\
\hspace*{1.1in}$\pm $0.0005  $\pm$ 0.000001\\


\section{Discussion}

\subsection{The distance problem}

Duerbeck et al. (1979) found a distance of d=450$\pm70$ pc.,
$A_{V}=0.27\pm0.13$ and $M_{V(max)}=-5.3\pm0.5$ (Section 1). The 
later value is very faint compared with typical values of novae 
(Warner 1986, 1995). Adopting these numbers and using the
current visual magnitude of the nova (m$_{V}\approx 16$ - Section 2), 
we can use the distance modulus equation (Allen 1976):

\begin{equation}
m_{V}=M_{V}-5+5log(d)+A_{V}
\end{equation}

and find $M_{V(1997)}=7.4\pm0.6$ for DN Gem. This brightness is relatively 
faint for old novae (Warner 1986; Naylor \& Somers 1997), and suggests a 
very low rate of mass transfer. Alternatively, the distance estimate might 
be wrong. In addition, when we use equation (7) of Retter \& Leibowitz 
(1998), with a typical white dwarf mass of 1$M_{\odot}$, we find that 
Duerbeck et al.'s values put Nova Geminorum well below the critical line 
for the thermal stability limit. This means, that as a probable non-magnetic 
CV (see arguments for this below), the system should have regular dwarf 
nova outbursts such as those observed in Nova V446 Her 1960 (Honeycutt, 
Robertson \& Turner, 1995; Honeycutt et al. 1998). We, therefore, estimate 
the distance to the system in a different manner - by the $t_{2}-M_{V}$ 
relation (Warner 1995). Using $t_{2V}\approx 16d$ (McLaughlin 1960), we 
obtain d=1600$\pm600$ pc. We adopt this value, which is much larger than 
the previous estimate, in this work. The corresponding absolute magnitude 
of DN Gem in outburst and in 1998 are $M_{V(max)}=-7.7\pm1.0$, and 
$M_{V_{current}}=4.5\pm1.0$. These values are almost identical with the 
estimates of Warner (1986) using the same method, but with a slightly 
different value of $t_{2}$. They are typical of absolute magnitudes of novae. 

\subsection{Identification of the photometric period}

We propose that the periodicity that prevails in the light curve of DN Gem 
through more than five years of photometric observations, P$\approx$3.1
h, is the orbital period of the underlying binary system. This suggestion 
is based mainly on the fact that this period was present in all yearly 
light curves, with no apparent change in its value, amplitude or shape. 
Typical orbital periods of nova systems range between $P_{orb}$=1.5 - 9 h 
with a peak around 4 h (Diaz \& Bruch 1997), so the proposed orbital 
period of DN Gem fits in the observed orbital period distribution of novae. 
Naturally, radial velocity measurements are required in order to 
confirm this suggestion.

\subsection{Classification of the system and identification of the 
variation - general remarks}

There are a few mechanisms that can generate such variation in old
novae. In many cases the various forms of accretion (i.e accretion
disc, accretion columns and bright spot) cause modulations in the light. 
Accretion plays an important role in the light curve of novae after only 
a few months to years after outbursts (Retter et al. 1997; Retter, 
Leibowitz \& Kovo-Kariti 1998) and certainly many decades afterwards 
(Bode \& Evans 1989). We observed fluctuations of up to 20-30$\%$ 
in the mean magnitude of adjacent nights presented in the light curve of 
DN Gem (Fig. 1). Such modulations in CVs are usually interpreted as 
fluctuations of the accretion source.

A major question is whether the white dwarf has a strong magnetic field 
(the accretion is maintained through accretion column/s) or whether 
the system is non-magnetic (the accretion stream forms an accretion 
disc around the white dwarf). We believe that based on the spectrum of 
DN Gem, which is characterized by a strong continuum and only weak 
absorption lines (Hummason 1938; Williams 1983), the first scenario can 
be eliminated, since AM Her systems have strong emission lines in their 
spectra (Warner 1995). The typical optical spectra, radiated from 
accretion discs, are characterized by a strong continuum, because the 
disc is usually optically thick in classical nova systems a few decades 
after their eruptions (Retter, Naylor \& Leibowitz 1999).

In the next few sections we examine a few models for the nova and for the 
variation, and analyse them in the light of our data. One option - an eclipse 
of one of the light sources of the system (for example the accretion 
disc), can be immediatedly ruled out from an inspection of the shape 
of the mean light curve (Fig. 3), which is sinusoidal and does not 
resemble an eclipse. 

\subsection{An intermediate polar model?}

Intermediate polars (for reviews see Patterson 1994; Hellier 1995, 1996, 
1998) are a subclass of CVs. In intermediate polar systems the rotation of 
the primary white dwarf is not synchronized with the orbital motion of the 
binary system, unlike the AM Her systems. The spin periods found in 
intermediate polars are much shorter than their orbital periods (Patterson 
1994; Hellier 1996), and range between 33 s in AE Aqr (Hellier 1996) to 
1.44 h in Nova V1425 Aql 1995 (Retter, Leibowitz \& Kovo-Kariti 1998). It is 
believed that in most intermediate polars the accretion is maintained most
of the time through an accretion disc (Patterson 1994; Hellier 1996).

The power spectrum of Nova Geminorum (Fig. 2) is inconsistent with a
significant second periodicity in addition to the main period (Section 3). 
Our relatively long (3-4 m) exposure times do not, however, allow us to
eliminate the possibility of a very short spin period of the order of a
few tens of seconds. Nevertheless, Robinson \& Nather (1977) made a
high speed photometric search in unfiltered light for very short
periods in a sample of ten novae, including DN Gem.  No such period was
found in Nova Gem 1912 up to 0.0022 mag and to the Nyquist frequency
(0.1 Hz). This fact suggests that unless the spin period of Nova Gem 1912
is shorter than all known intermediate polar periods, there is no short
spin period in Nova Geminorum.

The intermediate polar model cannot be definitely ruled out. We could
consider the observed 3.1 h period as the spin period itself,
speculating that a possible longer period (the orbital period) has not
been detected so far.  If this suggestion were true, the spin period
of this system would be the longest among all known intermediate polar 
systems. However, we believe that a better interpretation for the 3.1 
h period is that it is the orbital period of the binary system 
(Section 4.2).

We conclude that it is very unlikely that an intermediate polar model
fits the nova.

\subsection{A permanent superhump system?}

Permanent superhumps are quasi-periodic oscillations that appear in the
light curve of relatively bright CV systems (Osaki 1996; Retter \& 
Leibowitz 1998; Patterson 1999; Retter \& Naylor in preparation), with 
short orbital periods ($P_{orbital}$=17-225 m). The superhump period is 
a few per cent longer than the orbital period (la Dous 1993). The 
precession of the accretion disc, which surrounds the white dwarf is 
believed to be the reason for the light variation (Osaki 1989, 1996).

As a bright, short orbital period CV, Nova Gem 1912 is a natural permanent 
superhump candidate. The fact that the power spectrum of the nova (Fig. 2) 
does not reveal two periodicities in the light curve does not rule out 
this hypothesis. First, we believe that we can reject the possibility 
that the periodicity we detect in Nova Geminorum is a superhump period, 
based on the stability of the observed period (Sections 3 and 4.2) and on 
the fact that the superhump variation is believed to be stronger in the 
blue bands (Warner 1995).  Observationally, in one well-investigated object 
(Nova V1974 Cygni 1992), the amplitude of the permanent superhump variation 
was very similar in the $B$, $V$ and $I$ bands (Semeniuk et al. 1994, 1995; 
Olech et al. 1996; Skillman et al. 1997; Retter et al. 1997). This is 
unlike the variation in DN Gem, which is below our detection limit in the 
$B$ band (Section 3)

Even if we interpret the detected periodicity as the orbital period of the 
binary system, the permanent superhump scenario could be applied to the nova. 
Permanent superhumps, unlike the strict meaning of their name, are not 
permanently present in the light curves of systems bearing this name. An 
example is again V1974 Cyg, a well established permanent superhump system 
(Skillman et al. 1997; Retter et al. 1997), which is probably the best 
observed of its kind. The superhump variations were detected in 1994, and 
since then have been present most of the time in the nova light curve. During 
a few weeks in 1995 July-August, the superhump modulation either decreased 
sharply in amplitude or completely disappeared from the light curve (Retter et
al. in preparation).  Other permanent superhump systems show similar behaviour 
(Patterson 1998). We should also keep in mind that, as already mentioned in 
Section 2, most of the observations presented in this work, were not obtained 
in ideal conditions, but are rather low quality data, because they were 
carried out during bright phases of the moon. The resulting high noise level 
in the power spectrum (Fig.  2) may, therefore, hide a possible permanent 
superhump periodicity of low amplitude, or the orbital period itself, if the 
observed periodicity is interpreted as a superhump period.

We try here to test the permanent superhump hypothesis. Based on the 
tidal-disc instability model (Osaki 1989, 1996), and assuming that the
accretion disc is the dominant light source in the visual band, Retter
\& Leibowitz (1998) developed a way to check the thermal stability of
CVs. Permanent superhumps are supposed to be developed in thermally 
stable discs, while SU UMa systems (performing regular superhumps in their
light curves) are thermally unstable. In this method, 
the current mass transfer rate of the CV system is estimated by a few system 
parameters, and compared with the critical thermal instability value. 

We use the parameters of DN Gem in the equations of Retter \& Leibowitz. The 
current visual magnitude of DN Gem is $m_{V}$$\approx$16.0 (Section 2), and 
the distance, d=1600$\pm600$ pc (Section 4.1). We use the interstellar 
reddening, $A_{V}=0.27\pm0.13$, which was determined by Duerbeck et al. 
(1979), although it should be somewhat higher as we believe that they 
underestimated the distance (Section 4.1). To our knowledge, there is no 
cited value in the literature for the white dwarf mass, however 
$M_{WD}\approx 1 M_{\odot}$ is a typical value for nova primaries, and a 
large change in this value doesn't alter our final conclusion.  Inserting 
all these values into equation (8) of Retter \& Leibowitz (1998), we get
$\dot{M}\approx 4\pm3\times 10^{17}$g s$^{-1}$. The critical value of the 
mass transfer, which we get from equation (1) of Retter \& Leibowitz,
is about 2$\times 10^{17}$g s$^{-1}$, so the system is very close to the
thermal instability border-line. A definite conclusion cannot be made.

As a concluding remark for this section, we say, that unless a superhump 
periodicity is found in the light curve of Nova Gem 1912 in the future, 
in addition to the observed (orbital) period, this scenario cannot be 
applicable for the system and seems to be excluded by the observations.

\subsection{Light from the companion}

In this section we test the idea that the variation is generated by
light from the red dwarf. We discuss two models.

\subsubsection{An ellipsoidal variation?}

The ellipsoidal variation is caused by the distorted shape of the
secondary star in the binary system, due to the tidal
forces exerted by the primary white dwarf. The result is a change in
the surface area of the companion, seen by the observer at different
phases of the cycle. Such light curves typically have two maxima at phases
-0.25 and +0.25 relative to conjugation. We believe that there is only a
small chance that such a model fits the observational features of DN Gem, 
because a typical ellipsoidal light curve has a double a-symmetrical 
structure. The mean light curve of Nova Gem 1912 is sinusoidal (Fig. 3).  
When the mean light curve is folded onto the double period (6.14 h), there 
is no difference between the two minima within the error limit.  An ellipsoidal
variation model may, thus, fit the nova only if the change in the light
at the two different phases of the cycle mentioned above is
coincidentally similar to each other. 

A very strong argument againsts the ellipsoidal variation effect as an 
explanation for the observed periodicity, is that the secondary star 
is very weak, and thus doesn't contribute significantly to the overall
light curve of the nova. Using equation 2.102 of Warner (1995) with the
alleged period of 6.14 h (twice the periodicity), we find that the visual 
absolute magnitude of the red dwarf is only about $M_{V}\approx$ 8.0. 
This value, and the current absolute magnitude of DN Gem (Section 4.1) 
suggests that the secondary star contributes insignificantly to the light 
in the binary system - only about 4-10\%. Using the code we apply for
the irradiation effect in the next section, we find that the amplitude of
the ellipsoidal modulation can be only about 20\% of this value or less
than 2\% of the total flux. This value is inconsistent with the observed 
peak-to-trough amplitude of about 6\% (Section 3).

Based on the above two arguments, we conclude that it is unlikely 
that the variation is an ellipsoidal effect.

\subsubsection{An irradiation effect?}

The observed variation in Nova Gem can be simply explained by
light coming from the secondary star in the system, illuminated by the
vicinity of the white dwarf, or from the various forms of
accretion, i.e. an accretion disc, a bright spot, or an accretion
column. The typical shape of such a modulation is sinusoidal, which
fits the folded light curve of Nova Geminorum 1912 (Fig. 3).

We investigated the plausibility of the modulation originating from the
irradiation of the secondary star using the model described in Somers,
Mukai \& Naylor (1996) and Ioannou et al. (1999). The model consists of 
a steady-state accretion disc, flared so as to be triangular in cross 
section, with a hot central source which irradiates a Roche-lobe filling 
secondary star. We chose the white-dwarf to be 1 M$_{\odot}$, and the 
secondary star to be 0.34 M$_{\odot}$, appropriate for a main-sequence star 
which would fill the Roche-lobe at a binary period of 3.07 h (Bode \& Evans
1989). The chosen pole temperature was 3440$^{\circ}$K. The semi-opening 
angle of the disc was 10$^{\circ}$. The disc radius was set at 70\% of 
the Roche Lobe of the white dwarf - 0.7R$_{L_{1}}$, close to the tidal 
radius. For a flat disc, we then searched a two-dimensional $\chi ^2$ 
space in irradiating luminosity and binary inclination. At each point 
in the grid, we set the mass transfer rate to match the observed $I$-band 
magnitude, assuming three distances of 1.0, 1.6, 2.2 kpc (Section 4.1). 
Unsurprisingly the resulting $\chi ^2$ space showed that low inclinations 
required high irradiating luminosities, while high inclinations  - low 
luminosities. The resulting light curves are almost indistinguishable from 
sine waves, and hence are good fits to the data, until the inclination is 
greater than about 60$^{\circ}$, when the mutual eclipses of disc and 
secondary star begin to affect the shape. Thus, we can rule out these 
higher inclinations.

The irradiating luminosity is not outrageous for a flat disc (at a 
luminosity of about 10L$_{\odot}$), even at an inclination of only 
about 10$^{\circ}$. This is similar to the accretion luminosity, and 
therefore could be supplied even without a hot white dwarf. For a flared 
disc, the flare shadows the secondary star, and so higher luminosities are 
required. In this case, 10L$_{\odot}$ are required at 25$^{\circ}$.
Finally, we simulated light curves in the $B$-band for typical parameters
which fit the $I$-band data. These have modulations of around 2 per cent, 
which are consistent with the one $B$-band light curve we have (Section 3). 

In Fig. 4 we display an example of the constraints on the inclination 
angle, deduced from our model. The overall shape of the $\chi ^2$ space 
varies little with reasonable changes to the chosen parameters, although 
the position of the contour depends on the parameters chosen. The distance 
used was 1.6 kpc, which fixed the mass transfer rate by the requirement to 
match the overall flux level. The $\chi ^2$ minimum was rescaled to give a 
reduced $\chi ^2$ of 1, and the contour shown is the 90$\%$ confidence 
region for three parameters of interest (inclination, irradiating luminosity 
and mass transfer rate). High inclination angles are prevented because of 
the absence of an eclipse in the mean light curve. Low inclination angles 
are favoured only for very high luminosities of the irradiated source.
In this example we found 10$^{\circ}$$\la$ i $\la$ 60$^{\circ}$, however 
taking into account the permitted range of the relevant input parameters, 
we adopt the constraints: 10$^{\circ}$ $\la$ i $\la$ 65$^{\circ}$. This 
result is consistent with a previous estimate of i $\approx 50^{\circ}$ 
(Warner 1986).

\begin{figure}

\rotate[l]{\epsfxsize=60mm
\epsffile{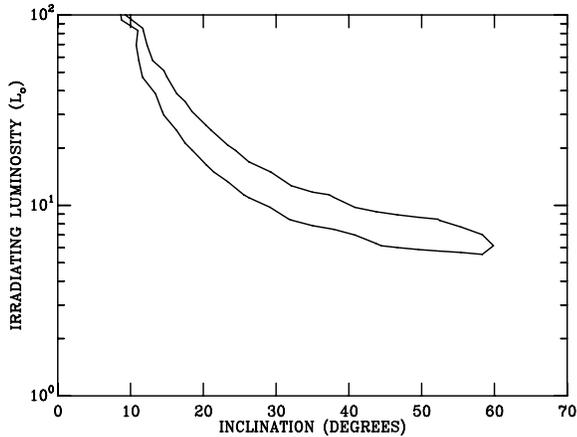} }

\caption{
An example $\chi ^2$ space for a 2-d grid search of inclination and 
irradiating luminosity, for the $I$-band data. The contour in the (i,L) 
plane defines the 90$\%$ confidence region. See text for more details.}

\end{figure}

In summary, an irradiated secondary star seems a likely source of the 
modulation. It reproduces the orbital light curve well, for reasonable 
irradiating fluxes, which could be provided either by a hot white dwarf, 
or by accretion luminosity.

\subsection{DN Gem as a nova inside the period gap}

The observed orbital period distribution of CVs presents a deficiency
of systems between about 2 h and 3 h, which is termed 'the period gap' 
(Warner 1995). Diaz \& Bruch (1997) used the updated data of Ritter \& 
Kolb (1998), and limited the gap to between 2.11 - 3.20 h. This interval 
encompasses the period of Nova Gem 1912. So far, the orbital periods of 
three other novae are inside the period gap - V Per 1891, QU Vul 1984 
and V2214 Oph 1988 (Wood, Abbott \& Shafter 1992; Baptista et al. 1993; 
Shafter et al. 1995; Diaz \& Bruch 1997). 

Our suggestion that the period observed in the light curve of DN Gem
is its orbital period (Section 4.2), makes DN Gem the fourth case of 
classical nova in the gap, as defined by Diaz \& Bruch 
(1997). The number of nova systems with orbital periods inside the 
period gap increases, then, to more than 10 per cent of the overall 
distribution (see Diaz \& Bruch 1997). It was suggested (Baptista et al. 
1993) that the period gap in the orbital period distribution of CVs does not 
exist for novae. The detection of the period in DN Geminorum supports this 
claim, however we will show below that it cannot be backed up statistically. 
Since there is no period gap for systems whose primaries harbour strong 
magnetic fields - AM Her systems (Warner 1995), Baptista et al. also argued 
that magnetic systems are favoured among novae with orbital periods that lie 
inside the period gap. Our findings (Section 4.3) don't support this idea. 

Undoubtedly the period distribution of novae is different from that of all
CVs, because there is a lack of nova systems below the period gap (Diaz \& 
Bruch 1997; Ritter \& Kolb 1998). We here subject to statistical test the 
idea of Baptista et al. (1993) that novae do not show a period gap at all.
The small number of systems below the gap make it hard to be quantitative 
about the period gap in novae. We, therefore, ask the question ``Is there 
a significant change in the number of systems per unit period interval at 
the upper end of the period gap, somewhere around 3 h?''. The answer to this 
question must be yes, if the gap exists, and so the question is interesting.
Furthermore, it turns out to be a well posed question in the statistical 
sense.

To answer this question we performed a one sided Kolmogorov-Smirnov test 
(Press et al. 1992) of the cumulative distribution of novae between 2.11 h 
(the bottom of the gap) and 6 h (arbitrarily chosen), against a model 
distribution with a constant number of systems per unit log period. This 
gave a 17 per cent chance that the two distributions arise from the same 
parent population, {\it i.e.} the existing data are still consistent with 
a smooth change of population density with period around the upper edge of 
the period gap. In other words, the data are consistent with no gap.

Obviously, with just four systems in the gap the statistics are rather poor,
so the next question we asked was if the data were consistent with a period
gap. To ask this question, we performed a two sided Kolmogorov-Smirnov test 
of the distribution of novae against all CVs (excluding novae) between 2.11 
and 6 h. Again, the distributions are consistent, with a probability of 75 per
cent of being from the same population. So the answer to the question ''Is 
there a period gap for novae?'' is as yet unsettled. There are simply too 
few systems to decide. 

Since primary stars in novae show a variety of magnetic behaviour (Warner 
1995), we should not, perhaps, expect that the distribution of classical 
novae will resemble that of AM Her systems, but rather be similar to that 
of all CVs.

In Fig. 5 we plot the discussed distribution. It is clear that the 
distribution of nova systems is not far removed from all CVs distribution, 
nor from a constant-period distribution.

\begin{figure}


\rotate[l]{\epsfxsize=60mm
\epsffile{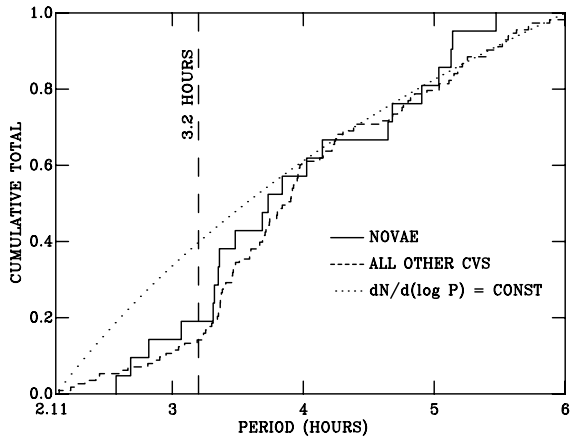} }

\caption{The accumulative distribution of novae, CVs,
and a theoretical model of constant distribution, versus the orbital period. 
The x-axis ranges are plotted from the lower end of the period gap until
an arbitrary chosen period of 6 h. The distribution of nova systems is
not significantly different from the other two distributions.
}

\end{figure}


\section{Summary}

The observed sinusoidal variation in DN Gem with the period $P\approx$ 
3.1 h is interpreted as the orbital period of the nova binary system. 
This suggestion places Nova Gem near the upper edge of the period gap. 
Three other novae have orbital periods in the period gap, as well. However, 
the rather poor statistics prevents a verification of the claim that there 
is no such gap for classical novae.

All possible explanations for the variation, we can think of, seem not to 
fit the observational data, except for the irradiation effect. We used this 
model to constrain the inclination angle - 
10$^{\circ}$ $\la$ i $\la$ 65$^{\circ}$.

\section{Acknowledgments}

We thank John Dan and the Wise Observatory staff for their expert assistance 
with the observations. We acknowledge Ohad Shemmer for measuring for us 
standard stars for the nova, and Coel Hellier for a fruitful discussion. 
A.R. is supported by PPARC. Astronomy at the Wise 
Observatory is supported by grants from the Israeli Academy of Sciences.


\begin{thebibliography}{99}

\bibitem{b9} Allen C.W., 1976, Astrophysical. Quantities, Athlone Press

\bibitem{b2} Baptista R., Yablonski F.J., Cieslinski D., Steiner J.E.,
1993, Ap.J., 406, L67

\bibitem{b4} Bode, M.F., Evans, A. 1989, Classical Novae, Alden Press, 
Oxford

\bibitem{b4} Diaz M.P., Bruch A., 1997, A\&A., 322, 807

\bibitem{b11} la Dous C., 1993, in Cataclysmic Variables \& Related Objects,
Hack M., la Dous C. (ed.), Centre National de la Recherce Scientifique,
Paris, France

\bibitem{b3} Duerbeck H.W., Lemke B., Willerding E., 1979, IBVS, 1711

\bibitem{b9} Efron B., Tibshirani R.J., 1993, "An Introduction to the
Bootstrap", Chapman \& Hall

\bibitem{b12} Hellier C., 1995, in Buckley D.A.H., Warner B., eds, 
Cape workshop on magnetic cataclysmic variables, ASP Conf.~Series, 85, 185

\bibitem{b12} Hellier C., 1996, in Evans N., Wood J.H., eds., Proc. IAU 
Colloq. 158, Kluwer, Dordrecht

\bibitem{b12} Hellier C., 1998, in Proc. of the 1996 Cospar Assembly,
Adv.~Space Res., 22, 973

\bibitem{b3} Honeycutt R.K., Robertson J.W., Turner G.W., 1995, Ap.J., 
446, 838

\bibitem{b3} Honeycutt R.K., Robertson J.W., Turner G.W., Henden A.A., 
1998, Ap.J., 495, 933

\bibitem{b3} Hummason M.L., 1938, Ap.J., 88, 228

\bibitem{b4} Ioannou Z. Naylor T., Welsh W.F., Catalan M.S., Worraker
W.J., James N.D., 1999, MNRAS, accepted

\bibitem{b4} McLaughlin D.B., 1960, in Greenstein L. (ed.), Stellar
Atmospheres, Vol. 4, 585

\bibitem{b11} Naylor T., Sommers M.W., 1997, in Wickramashinghe D.T.,
Bicknell G.V., Ferrario L., eds., IAU Colloq. 163, ASP Conference Series

\bibitem{b4} Netzer H. et al., 1996, MNRAS, 279, 429.

\bibitem{b10} Olech A., Semeniuk I., Kwast T., Pych W., DeYoung J.A.,
Schmidt R.E., Nalezyty M., 1996, Acta Astron., 46, 311

\bibitem{b3} Osaki Y., 1989,  PASJ, 41, 1005

\bibitem{b3} Osaki Y., 1996,  PASP, 108, 39

\bibitem{b4} Patterson J., 1994, PASP, 106, 697

\bibitem{b4} Patterson J., 1998, private communication

\bibitem{b4} Patterson J., 1999, in "Disk Instabilities in Close Binary 
Systems", eds. Mineshige S., Wheeler C., Universal Academy Press, in press

\bibitem{b4} Press W.H., Teukolsky S.A., Vetterling W.T., Flannery
B.P., 1992, 'Numerical Recipes', Cambridge University Press

\bibitem{b4} Retter A., Leibowitz E.M., 1996, IAU Circ., 6431

\bibitem{b4} Retter A., Leibowitz E.M., 1998, MNRAS, 297, L37

\bibitem{b4} Retter A., Leibowitz E.M., Ofek E.O., 1997, MNRAS, 283, 745

\bibitem{b1} Retter A., Leibowitz E.M., Kovo-Kariti O., 1998, MNRAS, 293, 145

\bibitem{b1} Retter A., Naylor, T., Leibowitz E.M., 1999, in "Disk 
Instabilities in Close Binary Systems", eds. Mineshige S., Wheeler C., 
Universal Academy Press, in press


\bibitem{b3} Ritter H., Kolb U., 1998, A\&A Suppl., 129, 83

\bibitem{b3} Robinson E.L. Nather R.E., 1977, PASP, 89, 572

\bibitem{b10} Semeniuk I., Pych W., Olech A., Ruszkowski M., 1994,
Acta Astron., 44, 277

\bibitem{b10} Semeniuk I., DeYoung J.A., Pych W., Olech A., Ruszkowski
M., Schmidt R.E., 1995, Acta Astron., 45, 365

\bibitem{b3} Shafter A.W.,, Misselt K.A. Szkody P., Politano M., 1995,
Ap.J., 448, L33

\bibitem{b3} Skillman D.R., Harvey D., Patterson J., Vanmunster T., 1997,
PASP, 109, 114

\bibitem{b3} Somers M.W., Mukai K., Naylor T., 1996, MNRAS, 278, 845

\bibitem{b3} Stetson P.B., 1987, PASP, 99, 191

\bibitem{b3} Warner B., 1986, MNRAS, 222, 11

\bibitem{b3} Warner B., 1995, Cataclysmic Variables Stars, Cambridge University Press

\bibitem{b3} Williams G., 1983, Ap.J.S., 53, 523

\bibitem{b3} Wood J.H, Abbott T.M.C., Shafter A.W., 1992, Ap.J., 393, 729

\end{thebibliography}
\end{document}